\documentclass{elsart}

 \usepackage{graphicx}
\usepackage{amssymb}
\newcommand{\ud}{\mathrm{d}}

\begin{document}

\begin{frontmatter}

% Title, authors and addresses

% use the thanksref command within \title, \author or \address for footnotes;
% use the corauthref command within \author for corresponding author footnotes;
% use the ead command for the email address,
% and the form \ead[url] for the home page:
% \title{Title\thanksref{label1}}
% \thanks[label1]{}
% \author{Name\corauthref{cor1}\thanksref{label2}}
% \ead{email address}
% \ead[url]{home page}
% \thanks[label2]{}
% \corauth[cor1]{}
% \address{Address\thanksref{label3}}
% \thanks[label3]{}

\title{Uniformity of the phase space and fluctuations in thermal equilibrium}

% use optional labels to link authors explicitly to addresses:
% \author[label1,label2]{}
% \address[label1]{}
% \address[label2]{}

\author[icm]{Arkadiusz Majka}\ead{majka@icm.edu.pl} and \author[icm,ipj]{Wojciech Wi\'slicki\corauthref{lab1}}\ead{wislicki@fuw.edu.pl}
\corauth[lab1]{Corresponding author}

\address[icm]{Interdisciplinary Centre for Mathematical and Computational Modelling, University of Warsaw, Pawi\'nskiego 5a, PL-02-160 Warsaw}
\address[ipj]{A. So\l tan Institute for Nuclear Studies, Ho\.za 69, PL-00-681 Warsaw}

\begin{abstract}
General relations are found between the measure of the uniformity of distributions on the phase space and the first moments and correlations of extensive variables for systems close to thermal equilibrium. 
The role played by the parameter of the Renyi entropy for the analysis of their fluctuations and correlations is studied.
Analytical results are verified and illustrated by direct simuations of quantum systems of ideal fermions and bosons.
Problems of finite statistics, usual in experiments and simulations, are adressed and discussed and solved by finding unbiased estimators for Renyi entropies and uniformities.
\end{abstract}

\end{frontmatter}

% main text

\section{Introduction}

Fluctuations and correlations in complex system provide unique information about its stability, scales and ranges of interactions between its constituents, and about qualitative changes in the system as a whole, as e.g. in phase transitions.
Studied at the microscopic level, such phenomena affect random variables on the phase space and, in particular, their local properties reflected in shapes or structure of singularities of their distributions.
On the other hand, they also influence global characteristics of the system, usually seen in moments of random variables.
Typical examples are relations between scales of correlations or strengths of intermittency and behaviour of thermodynamic potentials and quantities directly derivable from them, in particular thermal susceptibilities connected to variances of extensive variables, as specific heats or compressibilities.

The structure of the phase space can be studied using different formal tools.
Interesting from the physical point of view are links between mathematical properties of random variables on the phase space and observables for the system, either global, as e.g. moments of variables representing the system as a whole, or local, as characteristics of the phase space subsets related to subsystems.
In our previous work \cite{wislicki1} we found relations between equilibrium thermodynamics and the measure of the uniformity of the phase space, the latter being originally proposed in ref. \cite{beck1} in the context of multifractal sets and formulated in terms of generalized entropies or fractal dimensions.  In particular, we expressed the uniformity through thermodynamic potentials in the framework of the canonical and the grand canonical ensambles and proposed such approach for estimation of phase proportions in the vicinity of critical points.

In the following paper we develop the idea further and show how the uniformity of the phase space, being itself sensitive to local properties of the measure, can be used for direct calculation of first moments of such phenomenological global random variables as the total energy or the number of particles in the system.
We also show that the control parameter of the uniformity is at the same time the scaling parameter for intensive variables, as the temperature and the chemical potential, thus offering better understanding of relations between properties of complex systems at different scales.

Basic assumption for both the previous \cite{wislicki1} and the present works is that the system is in its stationary state, normally thermal equilibrium, or not far from it, which means that the probability densities are constant in time or quasistatic and the concept of temperature as a non-random parameter does make sense.
Apart from obvious interest in finding phenomenological interpretation for purely theoretical concept of the uniformity, our motivation for such effort is driven by experimental practice.
In many situations, one reasonably believes that stationarity conditions are fulfiled and probability densities can be estimated from data samples, but the equilibrium temperature of the system is unknown and hardly measurable directly.
Good examples for that are sets of remnants of nuclear disassembly processes in high energy collisions or that of stellar evolution where parts of final states are not detected or even not detectable at all because of experimental limitations, as hardly detectable neutrinos or objects out of aparatus acceptance.
Since the formalism can be applied to wider class of systems and phenomena, as e.g. to analyses of transportation networks or finance markets, for such cases parameters corresponding to the temperature or chemical potential do not even have clear operational interpretation and are not measurable just because are not observables {\it sensu stricto}.
As can be seen below, using uniformity one can predict the temperature evolution of the system from data taken at unknown temperature and eventually determine the temperature from the shape of the curve.
In addition, the uniformity function is parametrized by the real number which controls its sensitivity to different regions of the phase space, thus using experimentally available information in the most efficient way.

Following A. Renyi \cite{renyi11,renyi12} we introduce generalized entropies $I_q$, parametrized by the real number $q$
\begin{eqnarray}
I_q & = & \left\{\begin{array}{lc}
    \frac{1}{1-q}\ln \sum_{i=1}^{M(\Delta)} p_i^q, & \;\;\;\;\; q\neq 1 \\
    & \\
    -\sum_{i=1}^{M(\Delta)} p_i\ln p_i, & \;\;\; q=1
           \end{array}\right .
\label{one}
\end{eqnarray}
where the phase space is divided into $M(\Delta)$ numbered cells, $\{\Delta_i\}_{i=1}^{M(\Delta)}$, of the same volume $\Delta$ and
\begin{eqnarray}
p_i & = & \int_{\Delta_i} \ud\mu
\label{two}
\end{eqnarray}
$\ud\mu$ being the measure.
All sums run only over cells where $p_i\ne 0$.

Generalized Renyi dimensions are defined as
\begin{eqnarray}
D_q=-\lim_{\Delta\rightarrow 0}\frac{I_q}{\ln\Delta}.
\label{three}
\end{eqnarray}
All $D_q$s are positive and are monotonically decreasing functions of $q$.
By construction, the $I_q$ and $D_q$ are sensitive to non-uniformities of the measure on the phase space and $q$ may be used as a control parameter to specify the regions of the phase space one wants to monitor. 
In particular, for large positive $q$, only cells with the largest probabilities contribute to $I_q$ and $D_q$, thus making it sensitive only to the regions with highest event frequencies, and opposite to that, for large negative $q$, only regions with small probability density contribute.
In order to quantifiy the notion of uniformity and to standardize it, we use the Beck uniformity \cite{beck1}
\begin{eqnarray}
\gamma_q & = & \frac{D_q+q(q-1)D_q'}{D_q+(q-1)D_q'} \nonumber \\
         & = & \frac{I_q+q(q-1)I_q'}{I_q+(q-1)I_q'}
\label{four}
\end{eqnarray}
where $'$ stands for derivative over $q$.
For all $q$, $0\leq\gamma_q\leq 1$ and $\gamma_q=1$ and $\gamma_q=0$ correspond to the most uniform and the most non-uniform probability densities, respectively.
More detailed discussion of the properties of $\gamma_q$ can be found in refs \cite{beck1,wislicki1}.
It is worthwhile to note that $\gamma_q$ can be defined in eq. (\ref{four}) using both $I_q$ and $D_q$, as can be easily verified by following the line of argumentation leading to this final formula for $\gamma_q$ in ref. \cite{beck1}.
This is particularly important for finite statistics estimates of $\gamma_q$ and using it for description of local fluctuations, as discussed further in this paper.

\section{Equilibrium thermodynamics}

\subsection{The canonical ensemble}

Consider the canonical ensemble with the probability measure\footnote{In order to siplify formulae, we use hereon units with the Boltzmann and Planck constants equal to 1 and, in simulations for finite systems, particles' masses $m=1/2$ and unitary volumes.}
\begin{eqnarray}
\ud\mu=\frac{1}{Z(\beta)}e^{-\beta E(\vec r,\vec p)}\ud \vec r\,\ud\vec p
\end{eqnarray}
where $E$ stands for energy, $\beta=1/T$ for inverse temperature and $(\vec r,\vec p)$ for the phase space variables and 
\begin{eqnarray}
Z(\beta)=\int e^{-\beta E(\vec r,\vec p)}\ud \vec r\,\ud\vec p
\end{eqnarray}
is the partition function, where integration is taken over the whole phase space.
The Renyi entropy in this case is equal to
\begin{eqnarray}
I_q=\frac{1}{1-q}[\ln Z(q\beta)-q\ln Z(\beta)].
\end{eqnarray}
Using standard thermodynamics one finds a formula for specific heat at fixed volume
\begin{eqnarray}
c_V(q\beta)=q^2(1-q)I_q''-2q^2I_q'
\label{eight}
\end{eqnarray}
and explicit formula (\ref{one}) for $I_q$ leads to
\begin{eqnarray}
\frac{1}{q^2} c_V = \langle \ln^2 p\rangle_q - \langle\ln p\rangle_q^2
\label{nine}
\end{eqnarray}
where $p=\ud\mu/\ud\vec{r}\ud\vec{p}$.
By standard thermodynamics we mean the limit of $q\rightarrow 1$ where all quantities obtained from partition functions by using derivatives over $q\beta$ reduce to those known from classical thermodynamics. 
In particular, $c_V(q\beta)$ in eqns (\ref{eight},\ref{nine}) has the meaning of specific heat at temperature $q\beta$ and is defined using the first and the second derivatives of the partition function over $q\beta$. 
It is further related to the energy fluctuations at the temperature $q\beta$, as given in eq. (\ref{eleven}) below, and at $q\rightarrow 1$ one recovers the commonly known formula.
The average $\langle ..\rangle_q$ in eq. (\ref{nine}) is over so called escort distributions \cite{beck2}
\begin{eqnarray}
P_{q_i}=\frac{p_i^q}{\sum_{j=1}^M p_j^q}
\end{eqnarray}
where $p_i$s are given by eq. (\ref{two}).

Energy fluctuations are measured by the variance of energy
\begin{eqnarray}
\mbox{\em Var}[E(q\beta)]=\frac{1}{q^2\beta^2}c_V(q\beta).
\label{eleven}
\end{eqnarray}

From formulae above it is easy to reproduce obvious boundary case of zero fluctuations for delta-like, or microcanonical, probability distributions \\ $\ud\mu=\delta(E-E_0)\ud E$.
For $q=1$ one gets back usual thermodynamics, since $P_{1_i}=p_i$.

Noteworthily, for any function of temperature $A(\beta)$, the temperature scaling is equivalent to modification of averaging prescription
\begin{eqnarray}
\langle A(q\beta)\rangle = \langle A(\beta)\rangle_q.
\end{eqnarray}
This is understandable because any temperature dependence for macroscopic observables, being themselves mean values or functions of them, stays in the probability measure. 
The modification of the measure $p_i^q$ in case of Boltzmann like $p_i\sim exp(-\beta E)$ is just temperature rescaling.

Using escort averages and formulae for $\gamma_q$ and its derivative we get
\begin{eqnarray}
\gamma_q=q-\frac{\ln \sum_i p_i^q}{\langle\ln p\rangle_q}
\end{eqnarray}
and
\begin{eqnarray}
\gamma_q'=\Big(\frac{\langle\ln^2 p\rangle_q}{\langle\ln p\rangle_q^2}-1\Big)\ln \sum_i p_i^q
\end{eqnarray}
and $c_V$ (\ref{nine}) is given in terms of uniformity
\begin{eqnarray}
c_V(q\beta)=\frac{q^2(1-q)\gamma_q'}{(q-\gamma_q)^2}I_q
\end{eqnarray}
where
\begin{eqnarray}
I_q=\mbox{\em const}\cdot\exp\int_0^q \ud r \frac{(\gamma_r-1)}{(r-1)(r-\gamma_r)}.
\label{sixteen}
\end{eqnarray}

\subsection{The grand canonical ensemble}

For the grand canonical ensemble, the number of particles $N$ in the system is a random variable and the probability measure reads
\begin{eqnarray}
\ud \mu_N=\frac{1}{\Xi(\beta,\mu_c)}e^{\beta[\mu_cN-E_N(\vec r_N,\vec p_N)]}\ud\vec r_N\ud\vec p_N
\end{eqnarray}
where $\mu_c$ is the chemical potential and $E_N$ is the total energy of the $N$-particle state and $\Xi$ is the grand canonical partition function
\begin{eqnarray}
\Xi(\beta,\mu_c)=\sum_{N=0}^{\infty}\int e^{\beta[\mu_cN-E_N(\vec r_N,\vec p_N)]}\ud\vec r_N\ud\vec p_N 
\end{eqnarray}
where the integral is taken over the whole $N$-particle phase space $(\vec{r}_N,\vec{p}_N)$.
The Renyi entropy is equal to
\begin{eqnarray}
I_q=\frac{1}{1-q}[\ln\Xi(q\mu_c,q\beta)-q\ln\Xi(\mu_c,\beta)].
\end{eqnarray}
Standard thermodynamics gives an expression for the variance of $\beta E-\mu_c N$
\begin{eqnarray}
\mbox{\em Var}(\beta E-\mu_c N)(q\beta,q\mu_c)=(1-q)I_q''-2I_q'
\end{eqnarray}
where the total energy $E$ and the number of particles $N$ are in general correlated, so that
\begin{eqnarray}
\mbox{\em Var}(\beta E-\mu_c N)=\mbox{\em Var}(\beta E)+\mbox{\em Var}(\mu_c N)-2\;\mbox{\em cov}(\beta E,\mu_c N).
\end{eqnarray}

Similar reasoning as for the canonical ensemble gives
\begin{eqnarray}
\mbox{\em Var}(\beta E-\mu_c N)(q\beta,q\mu_c)=\langle\ln^2 p\rangle_q-\langle\ln p\rangle_q^2
\end{eqnarray}
where $p=\sum_N \ud \mu_N/\ud\vec{r}_N\ud\vec{p}_N$ and $\langle ..\rangle_q$ is the escort average over 
\begin{eqnarray}
P_{q_{i,N}}=\frac{p_{i,N}^q}{\sum_{j=1}^M\sum_{K=0}^{\infty}p_{j,K}^q}
\end{eqnarray}
and 
\begin{eqnarray}
p_{i,N}=\int_{\Delta_i}\ud \mu_N.
\end{eqnarray}

As for the canonical ensemble, for any function $A(\beta,\mu_c)$, the escort average is equivalent to the normal average for the temperature and the chemical potential scaled by $q$
\begin{eqnarray}
\langle A(q\beta,q\mu_c)\rangle=\langle A(\beta,\mu_c)\rangle_q.
\end{eqnarray}

Finally, the variance of $\beta E-\mu_c N$ can be expressed via $\gamma_q$
\begin{eqnarray}
\mbox{\em Var}(\beta E-\mu_c N)(q\beta,q\mu_c)=\frac{(1-q)\gamma_q'}{(q-\gamma_q)^2}I_q
\end{eqnarray}
where $I_q$ is given in terms of $\gamma_q$ by eq. (\ref{sixteen}).

It has to be noted that the meaning of the $q$ parameter, as used in present work, is not the same as for the Tsallis entropies (cf. e.g. ref. \cite{tsallis}) where it measures the degree of non-extensivity of that functions and may be further related to some properties of interactions in the system, as e.g. their effective ranges or, more generally, to the range of correlations.
Although for both the Renyi and Tsallis entropies the $q$ contributes via terms $p_i^q$, thus making them both sensitive to frequent or non-frequent events on the phase space, depending on the value of $q$, but the properties and physical meaning of them are quite different. 

\section{Local fluctuations and estimators of $D_q$, $I_q$ and $\gamma_q$}

For random variables represented by non-singular and smooth probability density functions, their Renyi dimensions do not depend on $q$ and the uniformity is equal to 1.
This is normally the case when the random variable represents total number of particles in the system.
But such approach is not suitable for description of local fluctuations.
In this case one has to consider densities being themselves random variables and dependent on the phase space random variables $\vec{X}_n$, as e.g. invariant density
\begin{eqnarray}
\rho_{\{\vec{X}_n\}_{n=1}^N}(\vec x)=\lim_{N\rightarrow\infty}\frac{1}{N}\sum_{n=1}^N \delta(\vec{x}-\vec{X}_n)
\end{eqnarray}
for which $D_q$ does exhibit non-trivial dependence on $q$.
Unfortunately, for finite statistics one cannot calculate directly an exact value of $D_q$ and has to use an estimator for it.
The usual method, known since at least two decades \cite{grassberger11,grassberger12}, is to plot $I_q$ as a function of $\ln\Delta$ (cf. eq. (\ref{three})) and determine the estimator $\widetilde{D}_q$ for $D_q$ from the slope.
Similar method was also used in ref. \cite{bialas1} for determination of non-Poisson fluctuations in rapidity distributions from factorial moments dependence on the bin size.
Such estimators are usually consistent but the problem of finite statistics is still there: for small cell size $\Delta$, where the estimate of $D_q$ is most reliable, we are out of statistics.

As mentioned above, the $\gamma_q$ can be calculated using either $D_q$ or $I_q$ (cf. eq. (\ref{four})).
Therefore, an alternative way is to determine the estimator $\widetilde{\gamma}_q$ for $\gamma_q$ defined in terms of $I_q$.
We managed to find an unbiased estimator $\widetilde{\gamma}_q$ using finite cell sizes and thus we are sure that finite statistics of experiment or simulation does not affect our measures in any systematic way.

It is commonly known that for the total number of events $N$, $n_i$ of them being in the $i$-th bin $\Delta_i$, the estimator $\widetilde{p}_i=n_i/N$ of $p_i$ is unbiased.
Formal proof is easy when $\widetilde{p}_i$ is represented as
\begin{eqnarray}
\widetilde{p}_i=\frac{1}{N}\sum_{j=1}^N \Theta_i(\vec{X}_j)
\label{twentyeight}
\end{eqnarray}
where $\Theta_i$ is an indicator function for the $i$-th bin, defined as
\begin{eqnarray}
\Theta_i(\vec{x}) & = & \left\{\begin{array}{lc}
    1, & \;\;\;\;\; \vec{x}\in \Delta_i \\
    0, & \;\;\; \mbox{\em otherwise}
           \end{array}\right .
\end{eqnarray}
For any integer $l\geq 1$, the unbiased estimator for $p_i^l$ is
\begin{eqnarray}
\widetilde{p^l}_i=\frac{n_i!}{(n_i-l)!}\frac{(N-l)!}{N!},\;\;\;\;\;\;\; n_i\geq l
\label{thirtyone}
\end{eqnarray}
since the expected value of $\widetilde{p^l}_i$ is
\begin{eqnarray}
E(\widetilde{p^l}_i)=\frac{(N-l)!}{N!}\sum_{j_1\neq\ldots \neq j_l}^N E[\Theta_i(X_{j_1})\ldots\Theta_i(X_{j_l})]
\label{thirtytwo}
\end{eqnarray}
and the number of non-zero terms, all of them equal to 1, at the righthand side of eq. (\ref{thirtytwo}) is equal to $n_i!/(n_i-l)!$.
In the limit of infinite statistics
\begin{eqnarray}
\lim_{N\rightarrow\infty}\widetilde{p^l}_i=\Big(\frac{n_i}{N}\Big)^l.
\label{thirtythree}
\end{eqnarray}
An obvious condition $n_i\geq l$ in eq. (\ref{thirtyone}) requires that statistics in each bin cannot be lower than the exponent $l$.
This limitation is even favourable from the numerical viewpoint, because for large $|q|$ the uncertainties of $\widetilde{p_i}$ propagate rapidly and so it is good to keep them limited.
If the condition $n_i\geq l$ is fulfiled then the relative statistical error on $p_i^l$ does not exceed $\sqrt{l}$.

This result can be extended to any real exponent $q$, requiring continuity of the estimator with respect to $q$ and using known properties of the gamma function, leading to the following estimator
\begin{eqnarray}
\widetilde{p^q}_i=\frac{n_i!}{N!}\prod_{k=1}^{N-n_i}(n+k-q).
\end{eqnarray}
Because of linearity of the expected value operation, any sum of terms $p_i^q$ is unbiased.

The $\gamma_q$ is an analytic function of $p_i$s $(i=1,\ldots,M(\Delta))$ and can be represented as a power series of terms $p_{i_1}^{q_1}p_{i_2}^{q_2}\cdot\ldots\cdot p_{i_M}^{q_M}$.
Since $p_i$s are independent of each other, except for marginal dependence from normalization, an expected value of the product factorize
\begin{eqnarray}
E\Big(\prod_{j=1}^M \widetilde{p_{i_j}^{q_j}}\Big)=\prod_{j=1}^M E\Big(\widetilde{p_{i_j}^{q_j}}\Big)
\end{eqnarray}
and
\begin{eqnarray}
\widetilde{\gamma}_q=\gamma_q\Big(\widetilde{p_{i_1}^{q_1}},\widetilde{p_{i_2}^{q_2}},\ldots,\widetilde{p_{i_M}^{q_M}}\Big)
\end{eqnarray}
is an unbiased estimator for $\gamma_q$. 
The only caveate can be raised for the limit of $M\rightarrow\infty$, but we do not go to zero with the cell size $\Delta$ and $M$ remains finite.
We find this feature of our estimator advantageous when one does not aim to estimate dimensions but only entropies, sufficient for finding uniformity, 
This completes our construction.

Some numerical studies of the statistical bias of $\gamma_q$ are given in the appendix.

\section{Simulations for ideal quantum systems}

We illustrate our theoretical results by applying them to grand canonical ensembles of ideal systems of fermions of spin 1/2 and bosons of spin 0.
Let $M$ denote the number of states of energies $\epsilon_r$, and $n_r$ are occupations of states $(r=1,\ldots,M)$, where $n_r=0,1$ for fermions and $n_r=0,1,2,\ldots $ for bosons.
Then $N=\sum_{r=1}^M n_r$ is the total number of particles in the system and $E_N=\sum_{r=1}^M \epsilon_r n_r$ is the total energy of the $N$-particle system.
The occupation numbers $\{n_r\}_{r=1}^M$ are the only random variables in this problem.

\subsection{The formulae}

Using notation
\begin{eqnarray}
x_r=e^{\beta(\mu_c-\epsilon_r)}
\end{eqnarray}
the grand partition functions are
\begin{eqnarray}
\ln\Xi & = & \left\{\begin{array}{lc}
             \sum_{r=1}^M \ln(1+x_r) & \;\;\;\;\; \mbox{\em fermions} \\
             -\sum_{r=1}^M \ln(1-x_r) & \;\;\;\;\; \mbox{\em bosons}
                    \end{array}\right . \\ \nonumber
\vspace{3mm}
       & = & \left\{\begin{array}{lc}
             -\sum_{r=1}^M\ln(1-\langle n_r\rangle) & \;\;\;\;\; \mbox{\em fermions} \\
             \sum_{r=1}^M\ln(1+\langle n_r\rangle) & \;\;\;\;\; \mbox{\em bosons}
                    \end{array}\right .
\end{eqnarray}
where 
\begin{eqnarray}
\langle n_r\rangle & = & \left\{\begin{array}{lc}
                         \frac{1}{x_r^{-1}+1} & \;\;\;\;\; \mbox{\em fermions} \\
                         \frac{1}{x_r^{-1}-1} & \;\;\;\;\; \mbox{\em bosons}
                                \end{array}\right .
\end{eqnarray}
and the probability distributions for occupations of states are equal to
\begin{eqnarray}
p(n_1,\ldots,n_M) & = & \left\{\begin{array}{lc}
    \frac{\prod_{r=1}^M x_r^{n_r}}{\prod_{r=1}^M(1+x_r)} & \;\;\;\;\; \mbox{\em fermions} \\
    \prod_{r=1}^M(1-x_r)x_r^{n_r} & \;\;\;\;\; \mbox{\em bosons}
           \end{array}\right .
\label{thirtynine}
\end{eqnarray}
Complete set of formulae for the mean, variance and covariance of $N$ and $E$, and for Renyi entropies and the uniformities is given below:
\begin{eqnarray}
\mbox{\em Var}(N) & = & \left\{\begin{array}{lc}
                        \langle N\rangle -\sum_{r=1}^M\langle n_r\rangle^2 & \;\;\;\;\; \mbox{\em fermions} \\
                        \langle N\rangle +\sum_{r=1}^M\langle n_r\rangle^2 & \;\;\;\;\; \mbox{\em bosons}
                               \end{array}\right .
\end{eqnarray} 
\begin{eqnarray}
\mbox{\em Var}(E) & = & \left\{\begin{array}{lc}
                        \sum_{r=1}^M(\langle n_r\rangle-\langle n_r\rangle^2) \epsilon_r^2 & \;\;\;\;\; \mbox{\em fermions} \\
                        \sum_{r=1}^M(\langle n_r\rangle+\langle n_r\rangle^2) \epsilon_r^2 & \;\;\;\;\; \mbox{\em bosons}
                               \end{array}\right .
\end{eqnarray}
\begin{eqnarray}
\mbox{\em cov}(N,E) & = & \left\{\begin{array}{lc}
                         \langle E\rangle - \sum_{r=1}^M\langle n_r\rangle^2\epsilon_r & \;\;\;\;\; \mbox{\em fermions} \\
                         \langle E\rangle + \sum_{r=1}^M\langle n_r\rangle^2\epsilon_r & \;\;\;\;\; \mbox{\em bosons}
                                \end{array}\right .
\end{eqnarray}
where $\langle N\rangle=\sum_{r=1}^M \langle n_r\rangle$ and $\langle E\rangle=\sum_{r=1}^M \langle n_r\rangle \epsilon_r$,
\begin{eqnarray}
I_q & = & \left\{\begin{array}{lc}
          -\frac{1}{q-1}\sum_{r=1}^M\ln[(1-\langle n_r\rangle)^q+\langle n_r\rangle^q] & \;\;\;\;\; \mbox{\em fermions} \\
          \frac{1}{q-1}\sum_{r=1}^M\ln[(1+\langle n_r\rangle)^q-\langle n_r\rangle^q] & \;\;\;\;\; \mbox{\em bosons}
                 \end{array}\right .
\end{eqnarray}
\begin{eqnarray}
I_0 & = & \left\{\begin{array}{lc}
          M\ln 2 & \;\;\;\;\; \mbox{\em fermions} \\
          \infty & \;\;\;\;\; \mbox{\em bosons}
                 \end{array}\right .
\end{eqnarray}
\begin{eqnarray}
I_{\infty} & = & \left\{\begin{array}{lc}
          -\sum_{r=1}^M \ln(1-\langle n_r\rangle) & \;\;\;\;\; \mbox{\em fermions} \\
          \sum_{r=1}^M \ln(1+\langle n_r\rangle) & \;\;\;\;\; \mbox{\em bosons}
                 \end{array}\right .
\end{eqnarray}
and
\begin{eqnarray}
\gamma_q & = & q-\frac{\sum_{r=1}^M\ln[(1-\langle n_r\rangle)^q-\langle n_r\rangle^q]}{\sum_{r=1}^M \frac{(1/\langle n_r\rangle-1)^q\ln(1-\langle n_r\rangle)+\ln\langle n_r\rangle}{(1/\langle n_r\rangle-1)^q+1}}  \;\;\;\;\; \mbox{\em fermions}
\label{fourtysix}
\end{eqnarray}
and
\begin{eqnarray}
\gamma_q & = & q-\frac{\sum_{r=1}^M\ln[(1+\langle n_r\rangle)^q-\langle n_r\rangle^q]}{\sum_{r=1}^M \frac{(1/\langle n_r\rangle+1)^q\ln(1+\langle n_r\rangle)-\ln\langle n_r\rangle}{(1/\langle n_r\rangle+1)^q-1}}  \;\;\;\;\; \mbox{\em bosons} 
\label{fourtyseven}
\end{eqnarray}

\subsection{Numerical results}

Numerical simulations for finite and bounded systems of ideal fermions and bosons were performed by choosing randomly occupations of states $n_r$ according to the probability distributions (\ref{thirtynine}).
Energies $\epsilon_r$ were determined from the wave equations for particles in a cubic box.
Simulations were performed for $q\in [0.01, 10]$ in steps of 0.15.
This means that each quantity, as variances of $E$, $N$, their covariances and uniformities, are calculated for each $q$. 
Then they are plotted, either as functions of $q$ in figs 1 and 2, or pairwise as $\gamma_q$ vs. Var($E$) etc., in figs 3 and 4.

In case of fermions the system consists of $N=10$ particles, the number of states $M$ is equal to 12 and the total number of configurations probed is 10,000. 

For bosons, the system contains $N=10$ particles, the number of states $M=4$ and the number of configuratons is equal to 10,000.
Simulations were performed for two temperatures, $T=0.6$ K and $T=3.6$ K. 
The temperature of Bose-Einstein condensation for this system is equal to 0.8 K.
 
In figs 1 and 2 the variances of the internal energy and the total numbers of particles and covariances of those two are plotted for two temperatures as functions of the temperature scaling parameter $q$.
Fig. 1 corresponds to fermions and fig. 2 to bosons.
One observes the variance of $N$ for bosons being decreasing function of temperature, unlike the other variances and covariances for bosons and fermions.
This is due to the specific temperature dependence of those quantities for bosons, with the $T^2$ term for $\mbox{Var}(E)$ and $\mbox{cov}(E,N)$ and no such term for $\mbox{Var}(N)$ (cf. e.g. ref. \cite{huang1}) and also to the fact that $T=0.6$ K is below and $T=3.6$ K is above the critical temperaturwe.

In figs 3 and 4 the unbiased estimators $\widehat{\gamma}_q$ of the uniformities $\gamma_q$ are shown as functions of the variances of $E$, $N$ and their covariance, for two temperatures. 
Fig. 3 refers to fermions and fig. 4 to bosons.
The maximas $\gamma_q=1$ correspond to $q=1$.

\section{Conclusions}

Our results illustrate general relations between characteristics of the probability measure on the phase space, viz. the uniformity of it, and thermodynamics of the system, thus providing with physical interpretation for the geometry of the event sets.
Their features were discussed in more detail for systems of fermions and bosons and, in particular, the behaviour of relevant quantities is illustrated by simulations in figs 1-4 and discussed in chapter 4.2.

Figs 1 and 2 illustrate how the temperature evolution of fluctuations of extensive quantities and their correlations can be predicted using the measurement at one temperature, corresponding to $q=1$ for each temperature.

As seen in figs 1-4, the shapes of curves are sensitive to the temperature.
In case of experimental data this allows for determination of the equilibrium temperature or, in general, the Lagrange coefficients of the problem, by fitting simulated curves.

An important advantage of the approach presented in this paper is that by increasing $q$ one probes those regions of the phase space where the probability density is most peaked. 
As seen in figs 1 and 2, fluctuations and correlations of extensive variables are decreasing functions of $q$. 
However, if the set of events under study is a mixture of signal and background events and if the background is distributed over the phase space more uniformly than the signal, which is often the case in experiments, the regions probed with high $q$ normally exhibit higher signal-to-noise ratio.
If the background is not separable from the signal on event-by-event basis, any quantity measured with high $q$, also fluctuations and correlations, may be less biased systematically.

In this work we found general relations between Renyi entropies, or Beck uniformities, and first moments and correlations of extensive variables for systems close to thermal equilibrium.
The role played by the degree parameter $q$ for the analysis of fluctuations and correlations was studied.
Our analytical results were illustrated by direct simulations for ideal quantum systems.
Problems of finite statistics in simulations or experiments were solved by finding unbiased estimators for expressions like $p_i^q$ and analytical functions of them, in particular for Renyi entropies and uniformities.

% The Appendices part is started with the command \appendix;
% appendix sections are then done as normal sections
% \appendix

% \section{}
% \label{}

\begin{figure}[h]
\begin{center}
\begin{tabular}{cc}
\includegraphics[width=70mm,height=65mm]{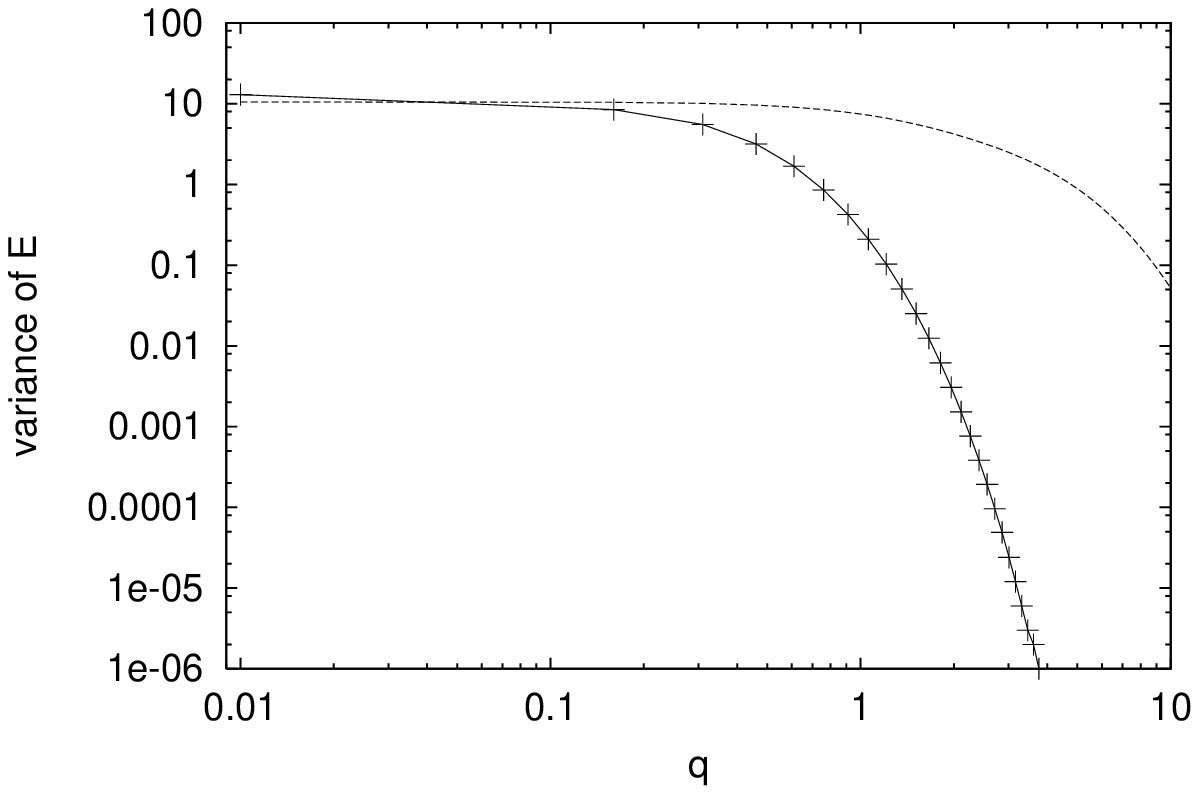} & \includegraphics[width=70mm,height=65mm]{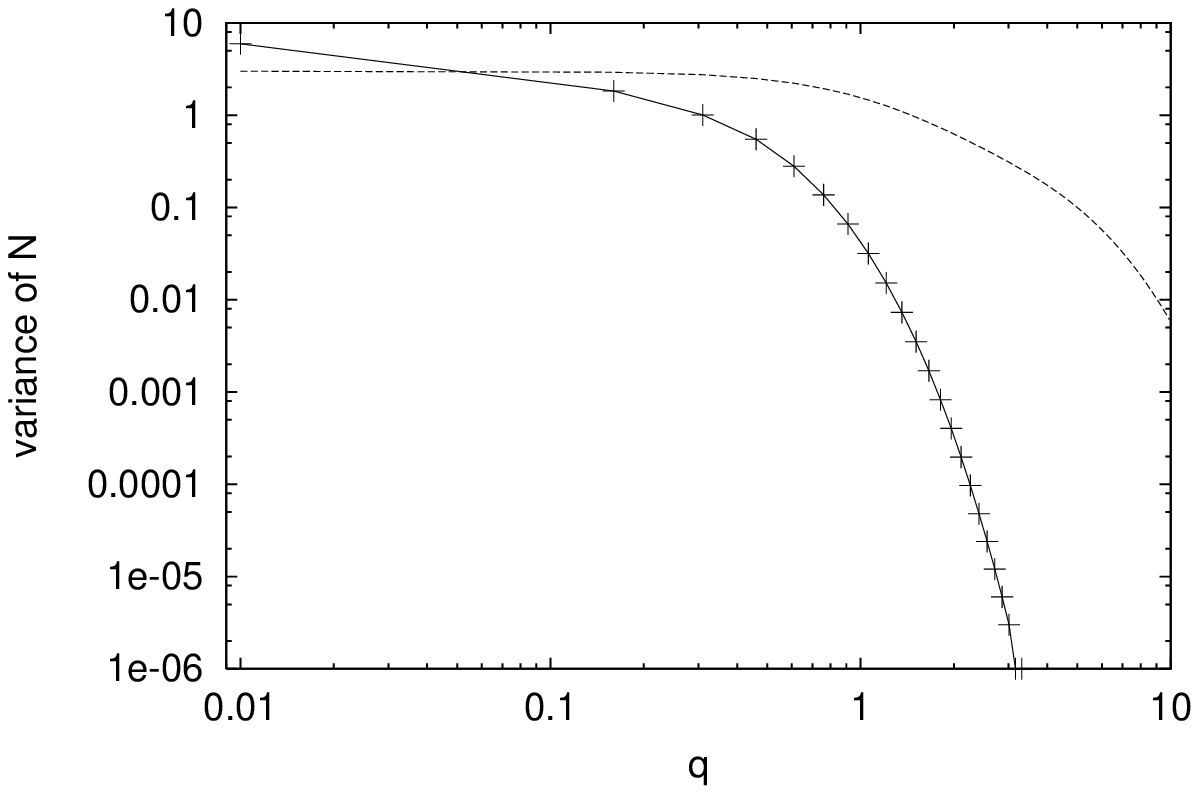} \\
\includegraphics[width=70mm,height=65mm]{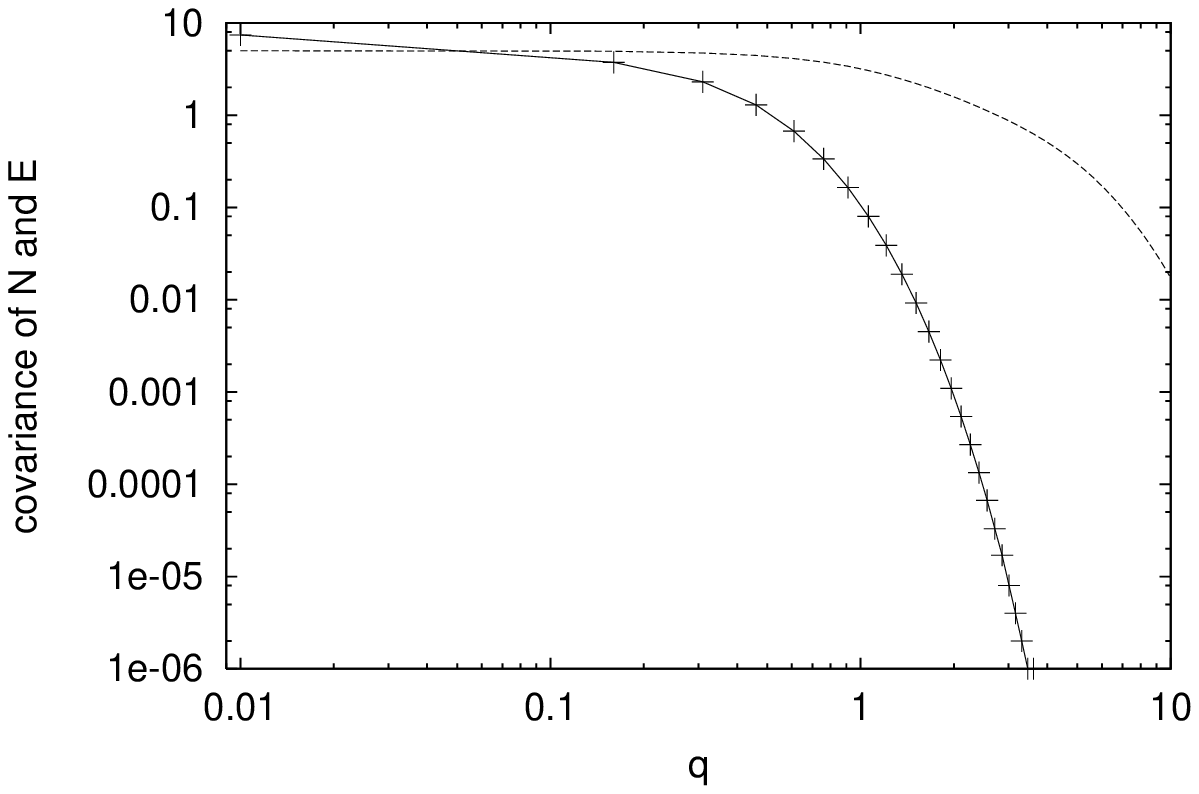} &
\end{tabular}
\end{center}
\caption{\em Results of numerical simulation for the ideal fermion gas: the variance of internal energy vs. $q$ (upper left), the variance of the total number of particles vs. q (upper right) and the covariance of internal energy and number of particles (lower). In all figures thicker curves with points correspond to $T=0.1$ K and the lighter ones to $T=1.1$ K.}
\end{figure}

\begin{figure}[h]
\begin{center}
\begin{tabular}{cc}
\includegraphics[width=70mm,height=65mm]{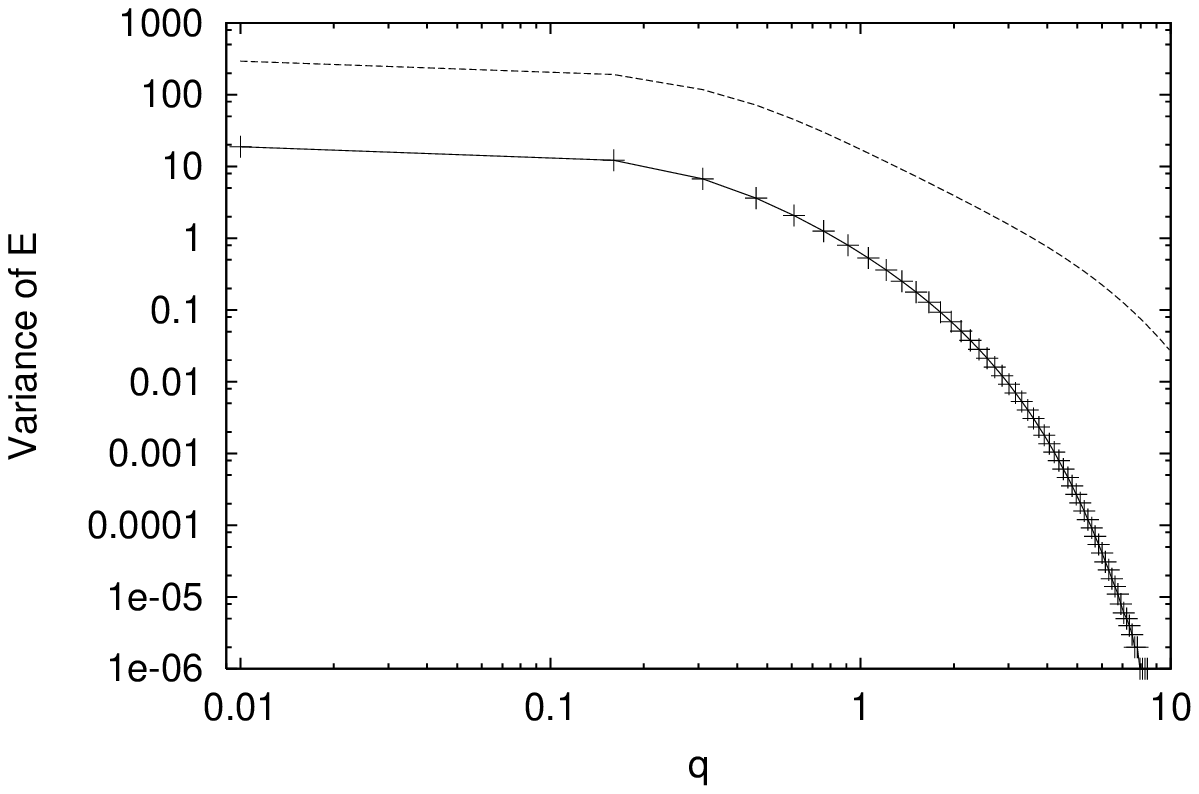} & \includegraphics[width=70mm,height=65mm]{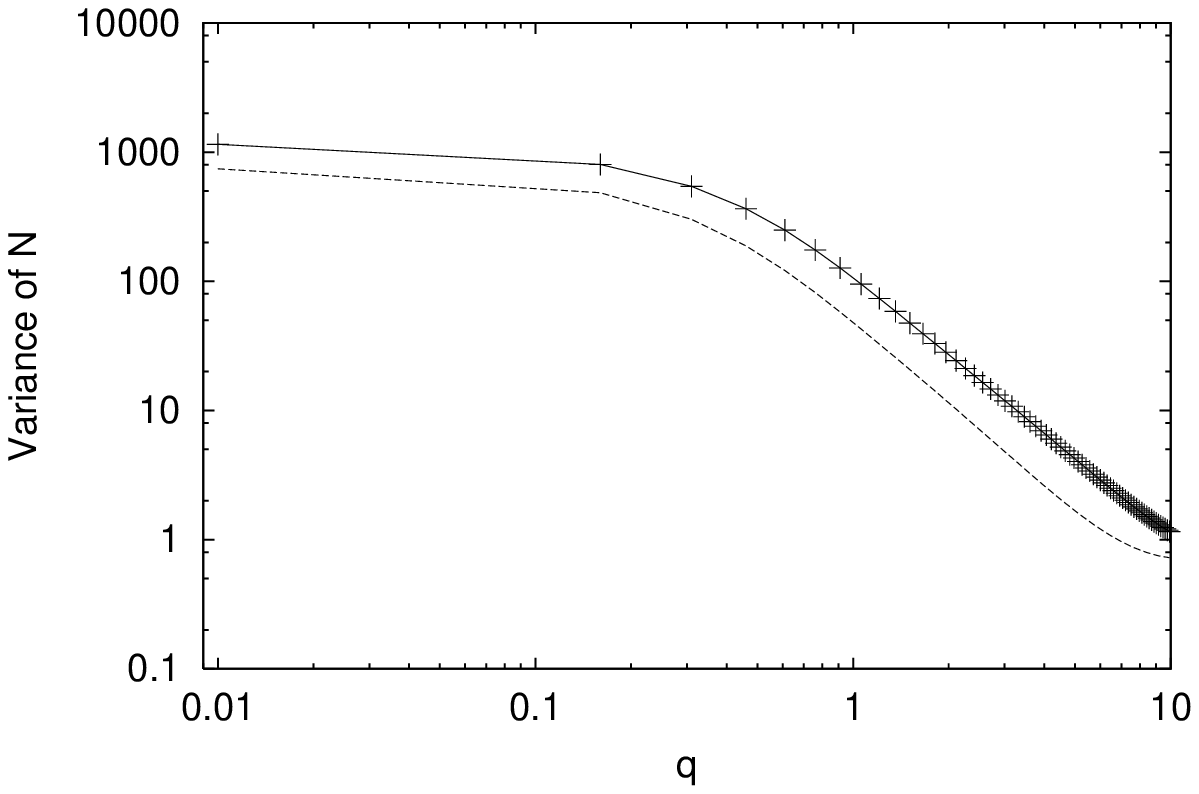} \\
\includegraphics[width=70mm,height=65mm]{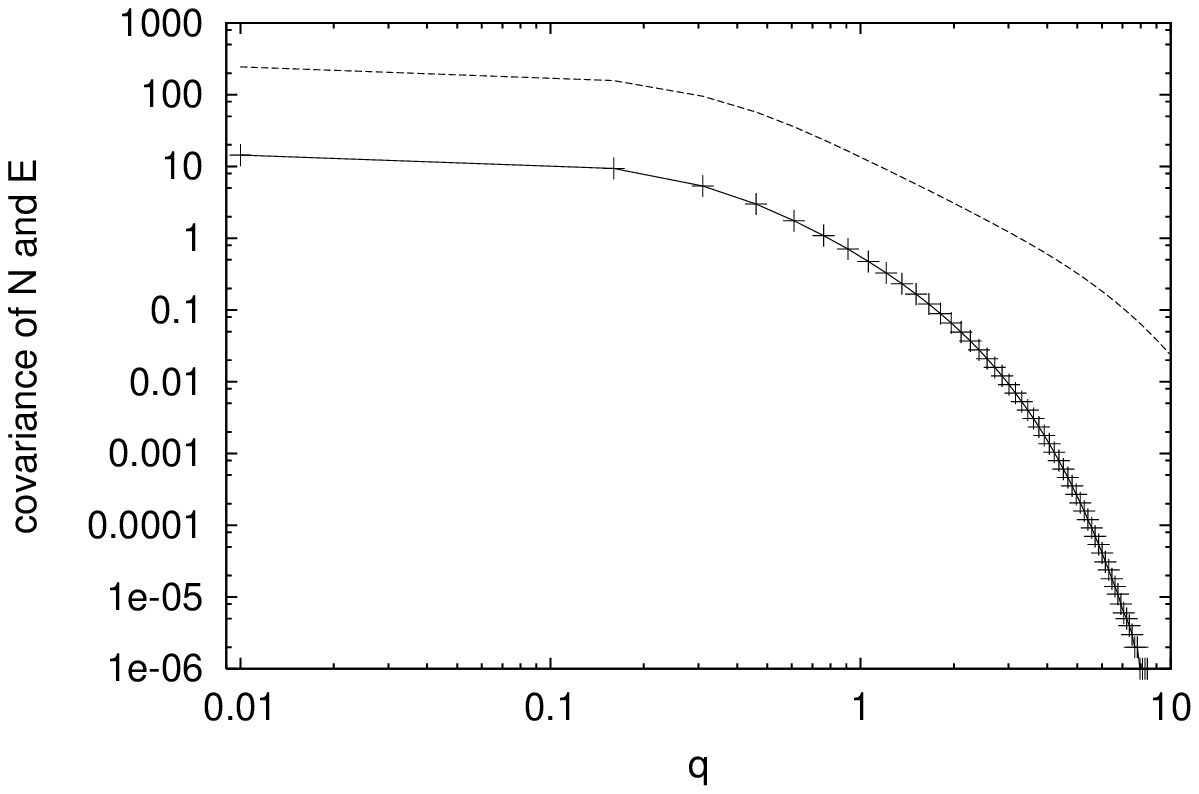} &
\end{tabular}
\end{center}
\caption{\em Results of numerical simulation for the ideal boson gas: the variance of internal energy vs. $q$ (upper left), the variance of the total number of particles vs. q (upper right) and the covariance of internal energy and number of particles (lower). In all figures thicker curves with points correspond to $T=0.6$ K and the lighter ones to $T=3.6$ K.}
\end{figure}

\begin{figure}[h]
\begin{center}
\begin{tabular}{cc}
\includegraphics[width=70mm,height=55mm]{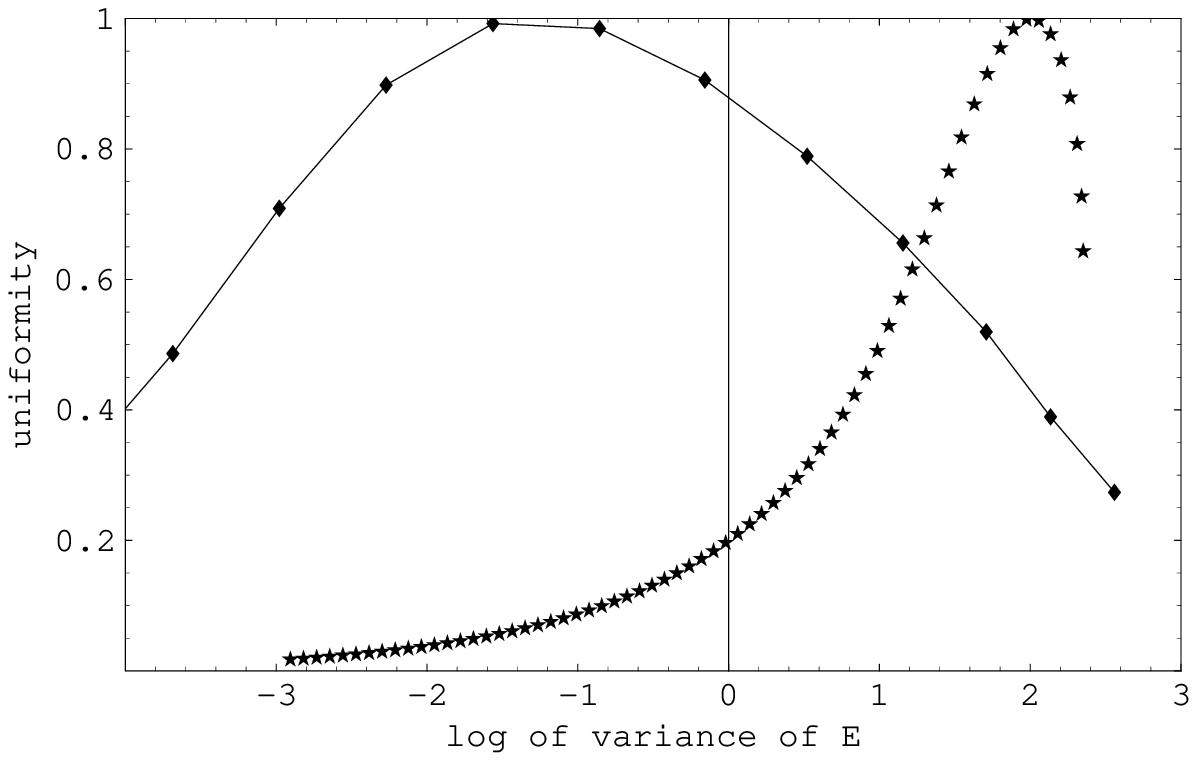} & \includegraphics[width=70mm,height=55mm]{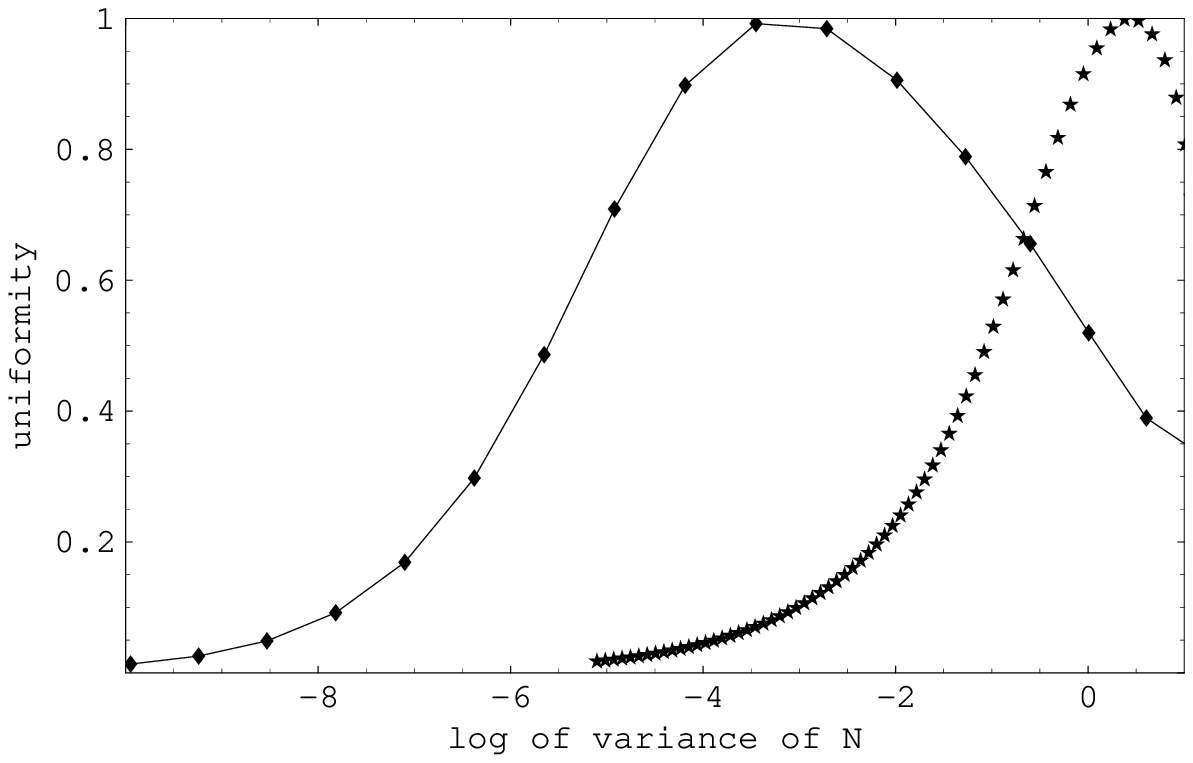} \\
\includegraphics[width=70mm,height=55mm]{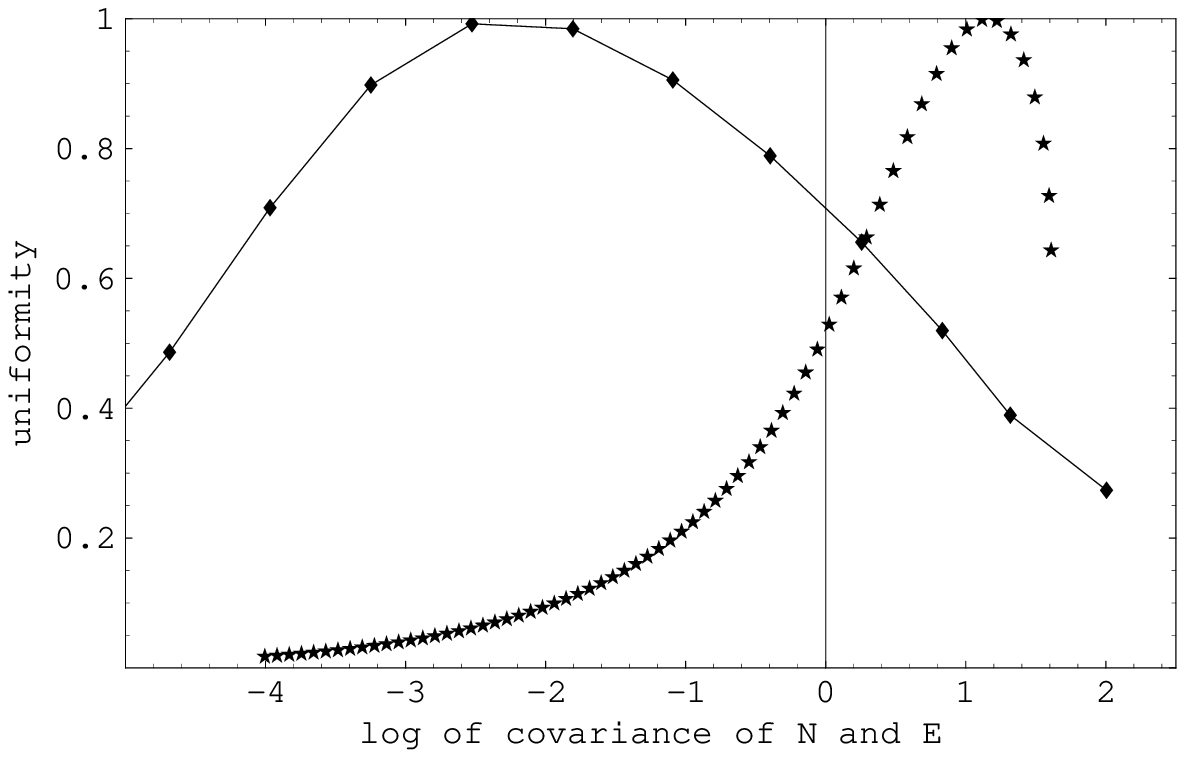} &
\end{tabular}
\end{center}
\caption{\em Results of numerical simulation for the ideal fermion gas: the uniformity vs. internal energy variance (upper left), the uniformity vs. total particles variance (upper right) and the uniformity vs. covariance of internal energy and number of particles (lower). Curves are drawn to guide ones eye. Continuous curves correspond to $T=0.1$ K and the dotted ones to $T=1.1$ K. For all uniformities their unbiased estimators are plotted.}
\end{figure}

\begin{figure}[h]
\begin{center}
\begin{tabular}{cc}
\includegraphics[width=70mm,height=55mm]{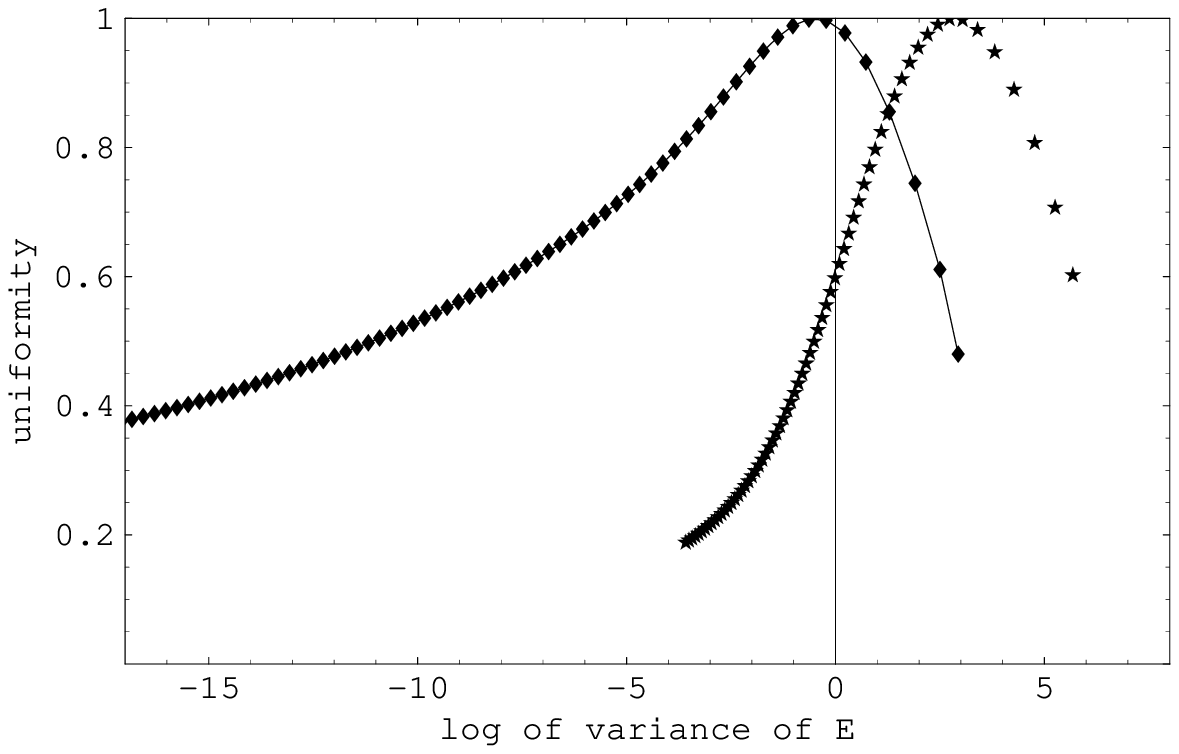} & \includegraphics[width=70mm,height=55mm]{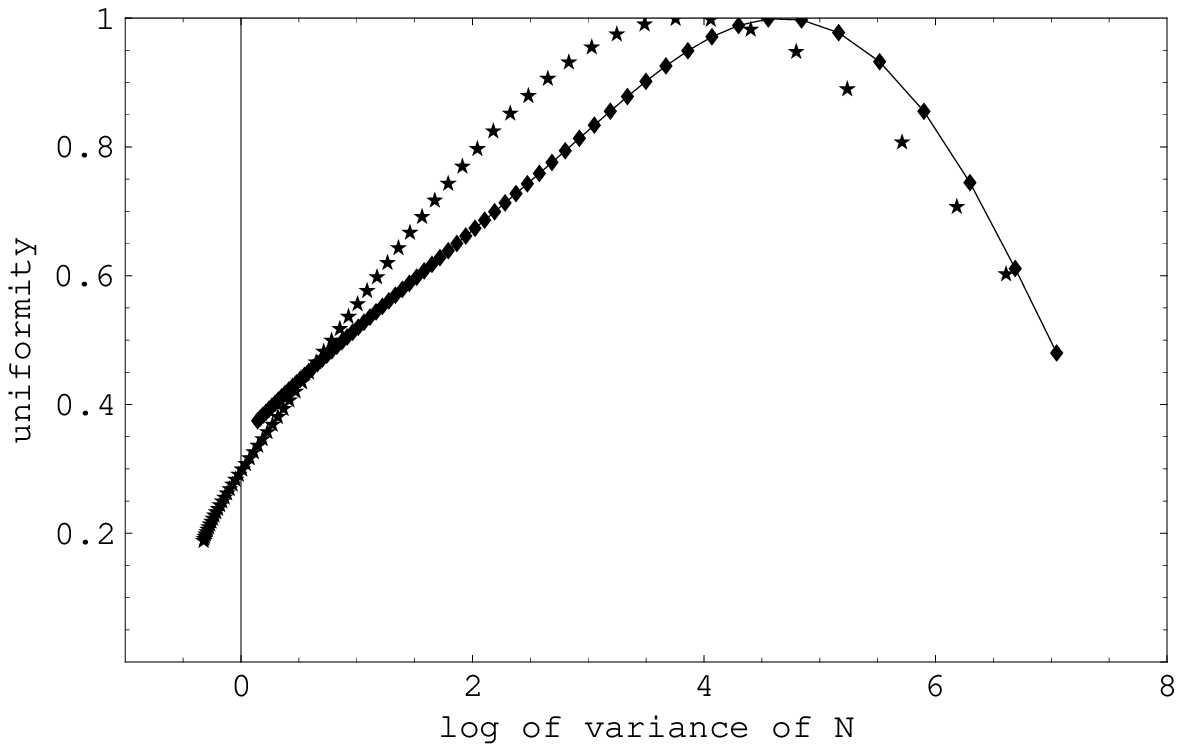} \\
\includegraphics[width=70mm,height=55mm]{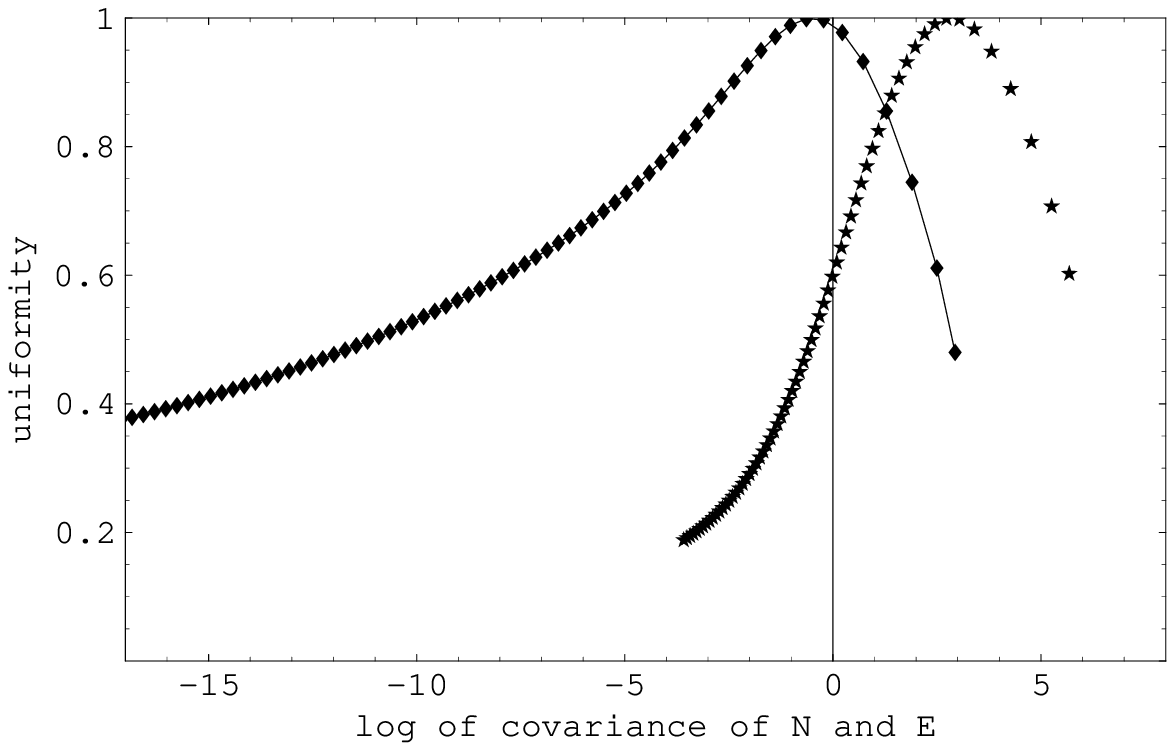} &
\end{tabular}
\end{center}
\caption{\em Results of numerical simulation for the ideal boson gas: the uniformity vs. internal energy variance (upper left), the uniformity vs. total particles variance (upper right) and the uniformity vs. covariance of internal energy and number of particles (lower). Curves are drawn to guide ones eye. Continuous curves correspond to $T=0.6$ K and the dotted ones to $T=3.6$ K. For all uniformities their unbiased estimators are plotted.}
\end{figure}

\appendix

\section{Notes on bias for estimators of $\gamma_q$ and numerical stability}

In order to understand effects of statistical bias for finite statistics, we performed comparative analysis of estimators of $\gamma_q$. The unbiased estimator $\widetilde{\gamma}_q$, as described in chapter 3, was compared to the biased estimator $\widehat{\gamma}_q$, calculated directly from formula (\ref{four}) using estimator $\widetilde{p_i}=n_i/N$ for $p_i$ (\ref{twentyeight}), but using estimator $\widehat{p_i^l}=\widetilde{p_i}^l$ for $p_i^l$ , and not that given by eq. (\ref{thirtyone}).
In figs A.1 and A.2 we present the ratio $R=\widehat{\gamma}_q/\widetilde{\gamma}_q-1$ which illustrates the bias of the estimator $\widehat{\gamma}_q$ for our simulations for fermions and bosons.

As seen in all figures, in the limit of $q\rightarrow 0$, $R$ is large negative which corresponds to strongly underestimated $\widehat{\gamma}_q$. 
Also, the bias becomes more pronounced for higher $q$, as compared to $0.5\lesssim q\lesssim 3$.
Such behaviour is due to the fact that for both very large and very small $q$, the $\gamma_q$ is smaller than for $q\simeq 1$ and there is more modestly populated phase space cells.
These cells stronger contribute to the bias because $(n_i/N)^l$ well approximates $\widetilde{p_i^l}$ only for high statistics (cf. eq. (\ref{thirtythree})).

In all cases additional checks were performed whether effects we observe are not due to numerical uncertainties which can be significant for $|q|\gg 1$.
In order to minimize numerical effects we made our calculations with accuracy of 70 digit numbers, using {\it Mathematica} software \cite{mathematica}. 
We found that unbiased estimators $\widetilde{\gamma}_q$ are in good agreement with analytic results (\ref{fourtysix},\ref{fourtyseven}).
This convince us of reliability of our simulation and smallness of marginal bias to the unbiased estimator $\widetilde{\gamma}_q$ which might eventually have come from the constraint $p_1+\ldots +p_M=1$, making $p_i$s not perfectly independent, as mentioned in chapter 3.
An exception is the case of high $q$ for fermions at $T=0.1$ K.
This case exhibits numerical instability, seen as intermittency for high $q$ in the upper left panel of fig. A.1.
This happens for both estimators and is due to relatively large number of poorly populated phase space cells, and thus small probabilities for them, which substantially contribute to strong propagation of rounding errors for high $q$.
For higher temperatures the uniformities are usually higher, there are less terms with small $p_i$s and such effects tend to disappear.

\begin{figure}[h]
\begin{center}
\begin{tabular}{cc}
\includegraphics[width=70mm,height=65mm]{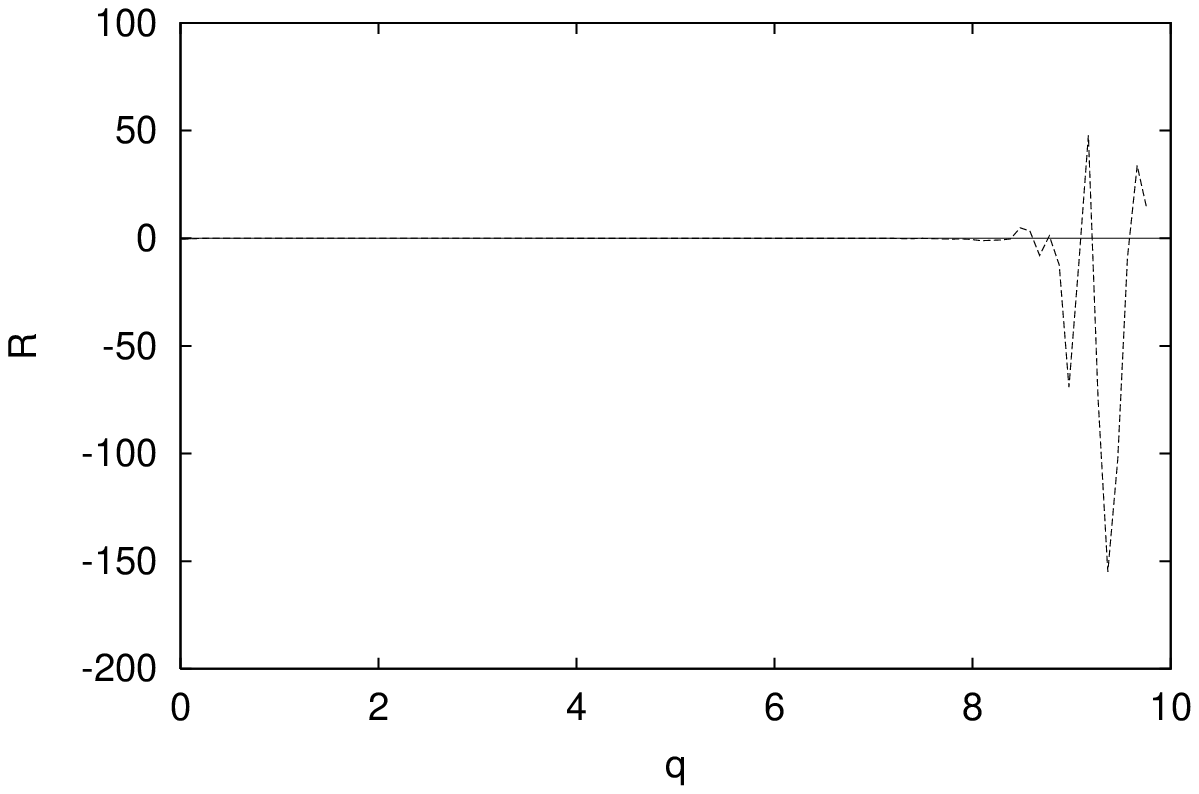} & \includegraphics[width=70mm,height=65mm]{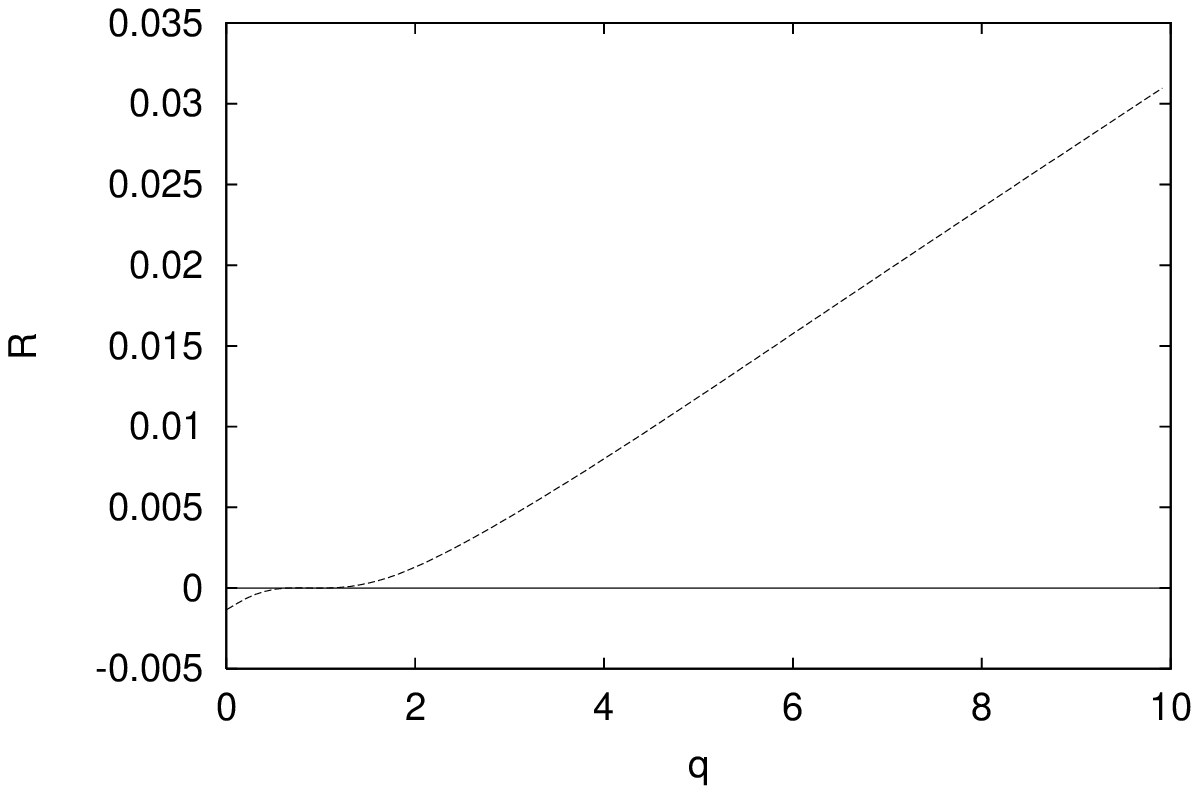} \\
\includegraphics[width=70mm,height=65mm]{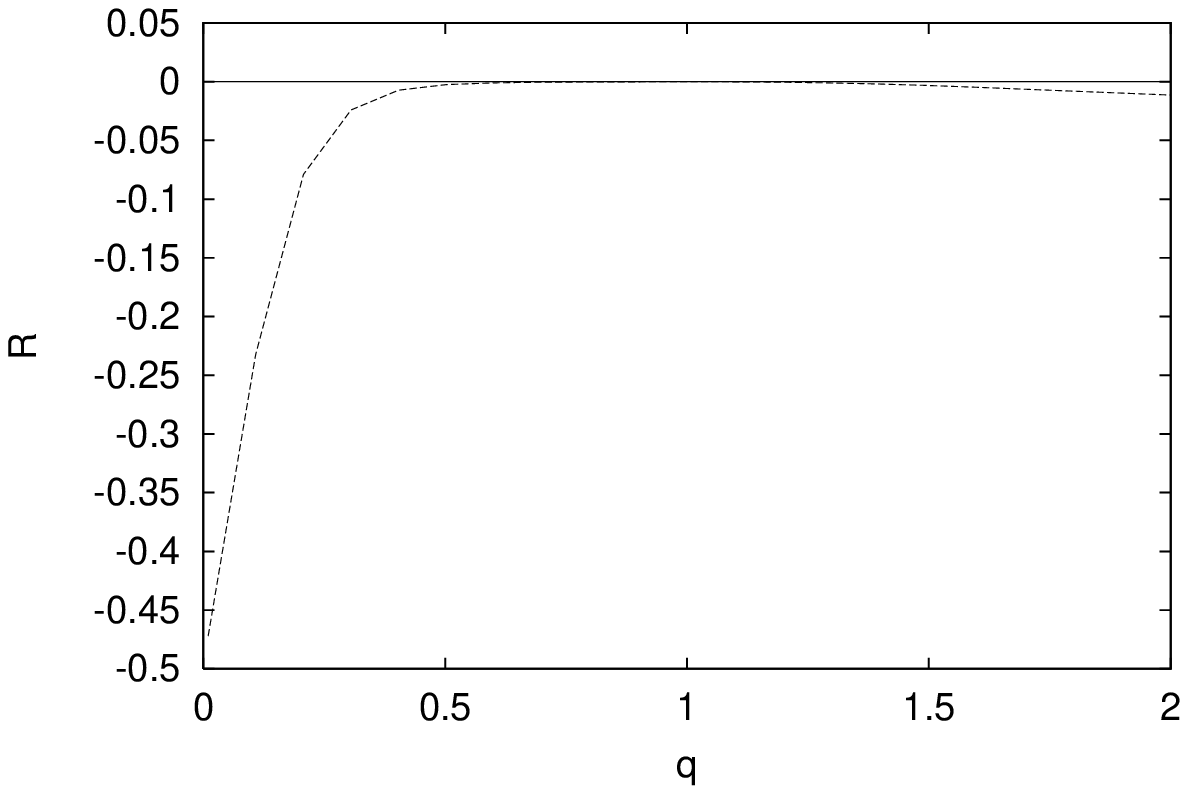} & \includegraphics[width=70mm,height=65mm]{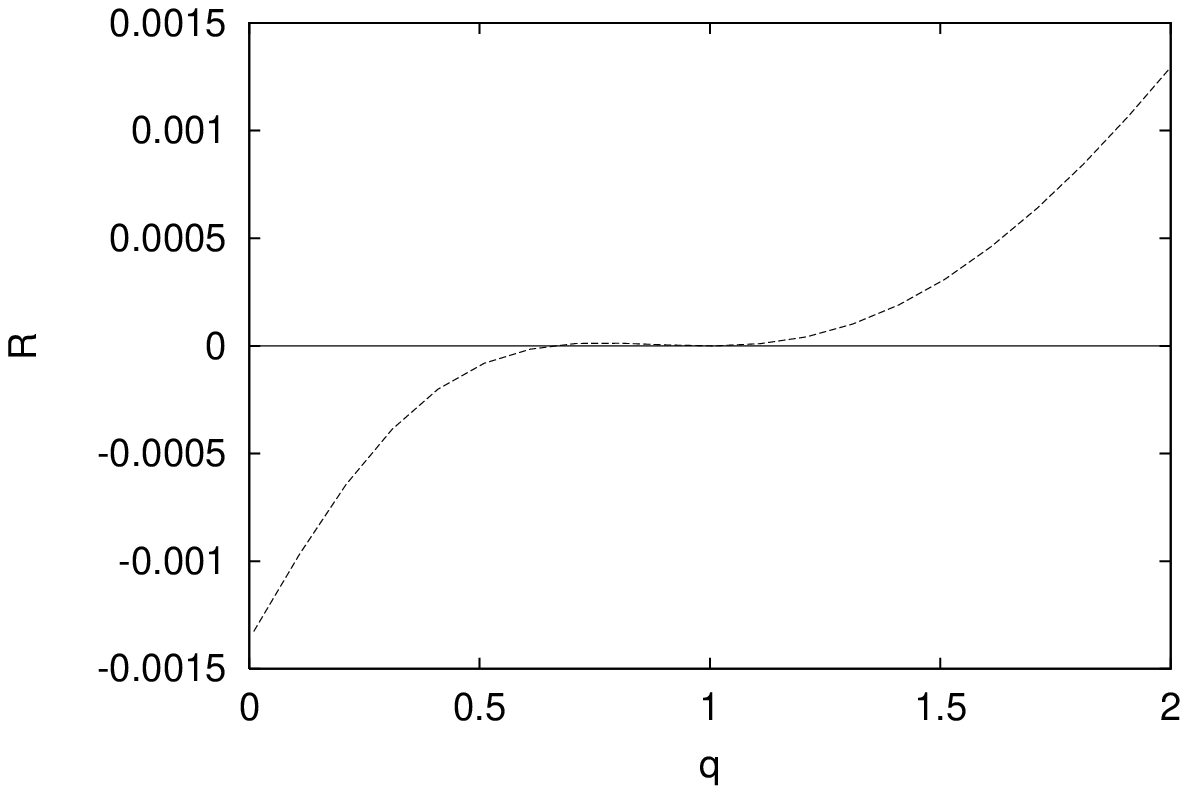}
\end{tabular}
\end{center}
\caption{\em The bias $R=\widehat{\gamma}_q/\widetilde{\gamma}_q-1$ for the uniformity for the ideal fermion system. The upper left panel corresponds to $T=0.1$ K, the upper right to $T=1.1$ K and the lower left and right panels refer to the same temperatures as upper ones but show only the low $q$ domain.}
\end{figure}

\begin{figure}[h]
\begin{center}
\begin{tabular}{cc}
\includegraphics[width=70mm,height=65mm]{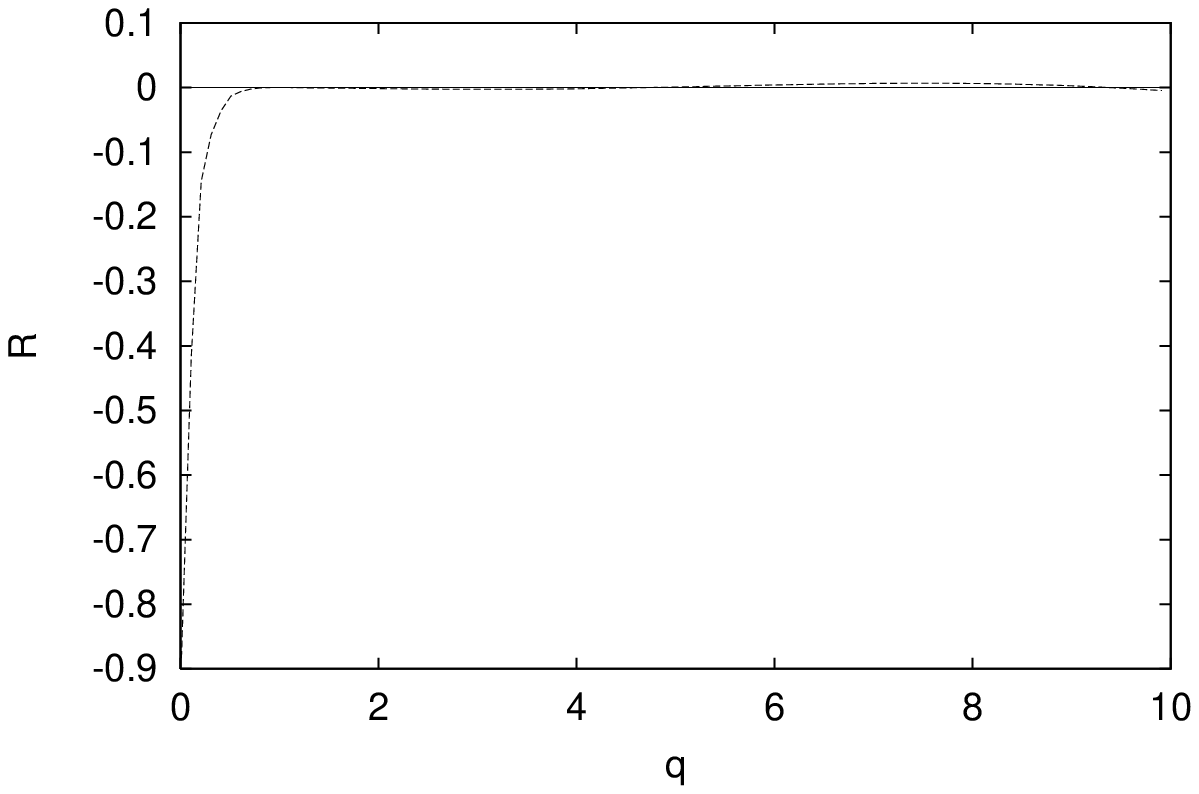} & \includegraphics[width=70mm,height=65mm]{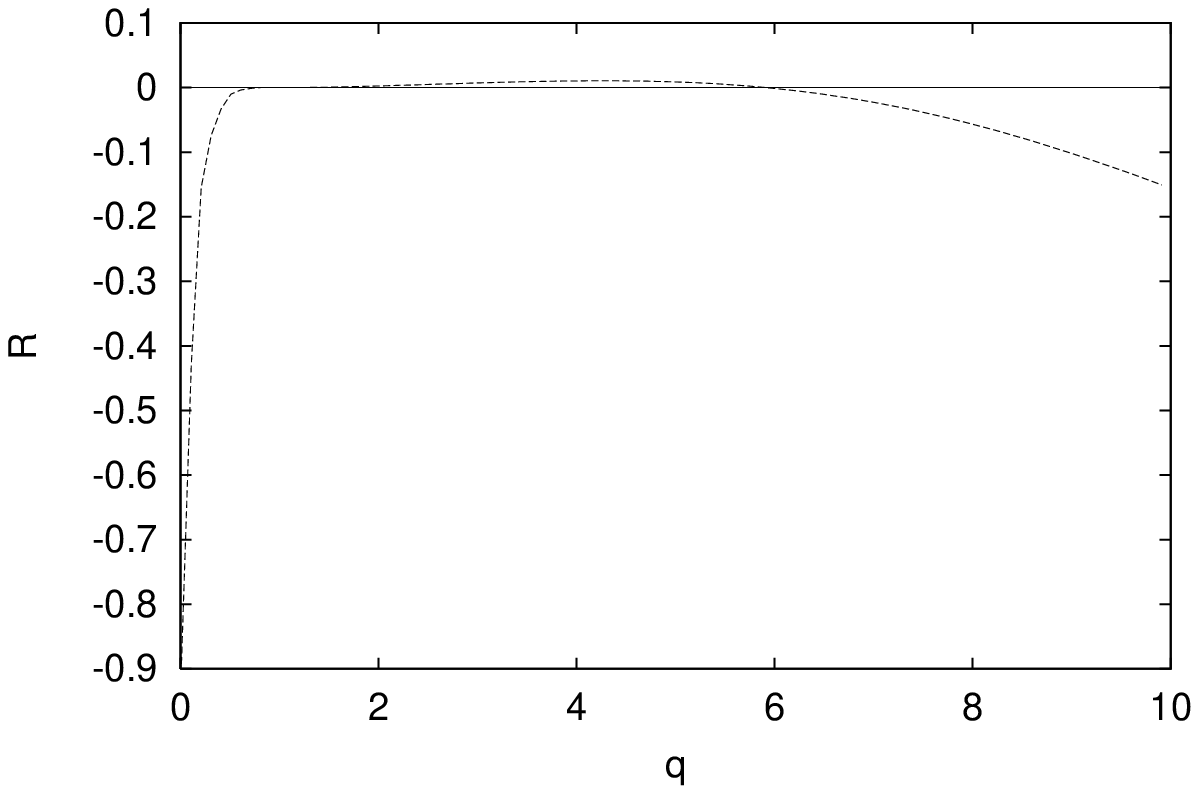} \\
\includegraphics[width=70mm,height=65mm]{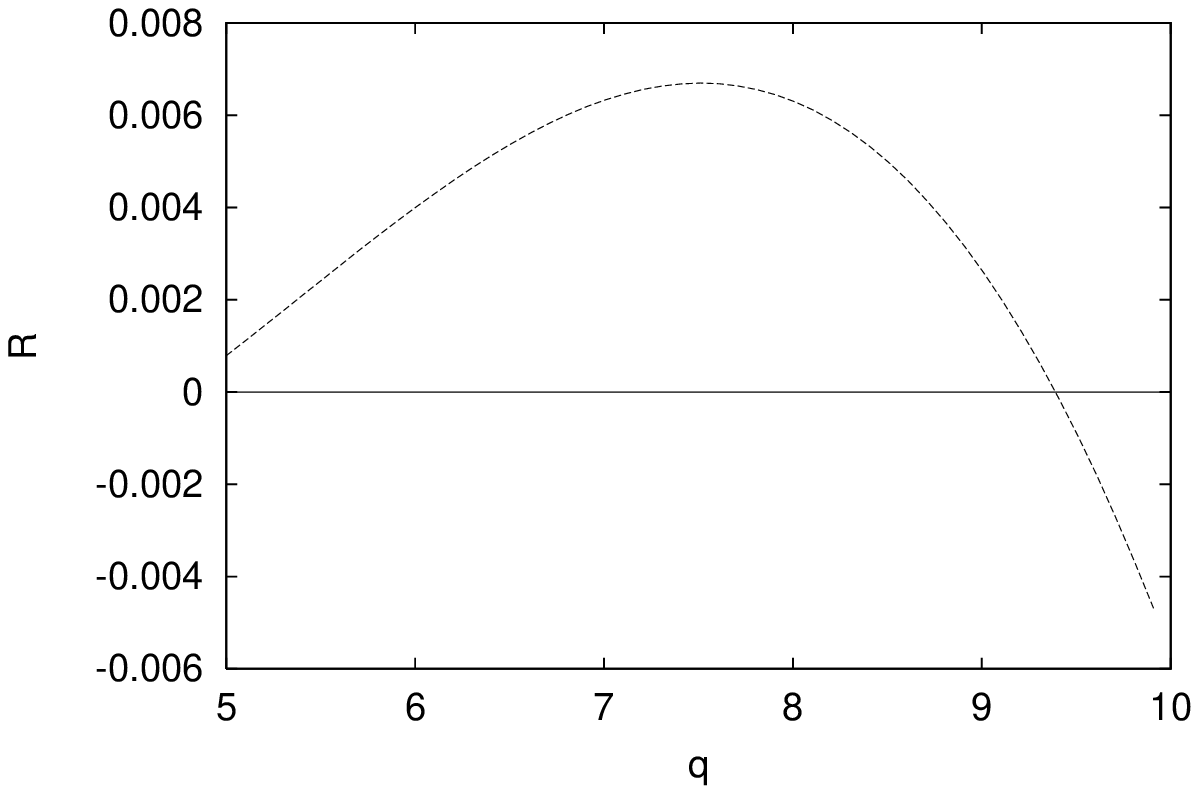} & \includegraphics[width=70mm,height=65mm]{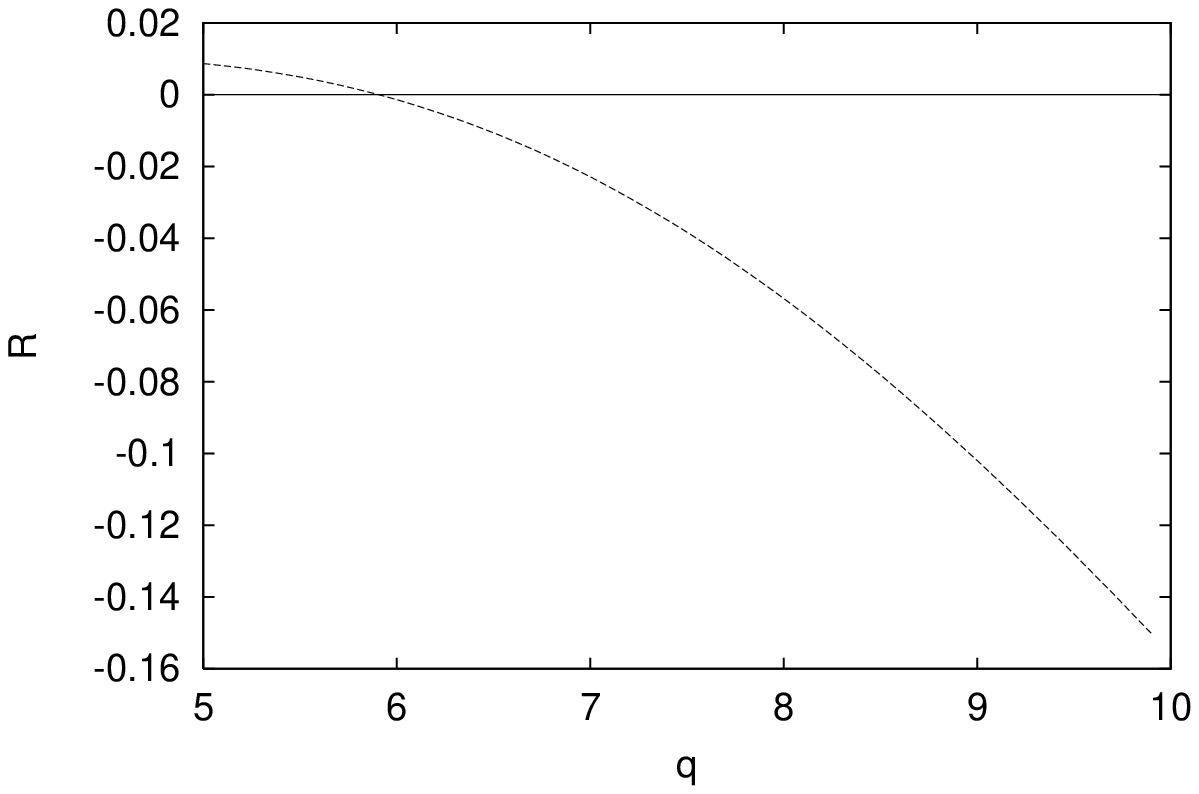}
\end{tabular}           
\end{center}                                                                                              \caption{\em The bias $R=\widehat{\gamma}_q/\widetilde{\gamma}_q-1$ for the uniformity for the ideal boson system. The upper left panel corresponds to $T=0.6$ K, the upper right to $T=3.6$ K and the lower left and right panels refer to the same temperatures as upper ones but show only the high $q$ domain.}
\end{figure}

\end{document}